\documentclass[aps,prd,preprint,amsmath,superscriptaddress,floatfix]{revtex4}

\usepackage{graphicx}

\begin{document}

\title{Testing Lorentz Invariance with Ultra High Energy Cosmic Ray Spectrum }

\author{ Xiao-Jun Bi$^{1,2}$, Zhen Cao$^1$, Ye Li$^1$, Qiang Yuan}

\affiliation{Key Laboratory of Particle Astrophysics, Institute of
High Energy
Physics, Chinese Academy of Sciences, Beijing 100049, 
People's republic of China\\
$^2$Center for High Energy Physics, Peking University, Beijing
100871, People's Republic of China }

\begin{abstract}
The Greisen-Zatsepin-Kuzmin cutoff (GZK cutoff) predicted at the Ultra High Energy Cosmic Ray
(UHECR) spectrum has been observed by the HiRes and Auger experiments.
The results put severe constraints on the effect of Lorentz
Invariance Violation (LIV) which has been introduced to explain the
absence of the GZK cutoff indicated in the AGASA data. Assuming
homogeneous source distribution with a single power-law spectrum,
we calculate the spectrum observed on the Earth by taking
photopion production, $e^+e^-$ pair production, and
the adiabatic energy loss into account. The effect of LIV is also
taken into account in the calculation. By fitting the HiRes monocular
spectra and the Auger combined spectra, we show that the LIV parameter
is constrained to $\xi=-0.8^{+3.2}_{-0.5}\times10^{-23}$ and
$0.0^{+1.0}_{-0.4}\times10^{-23}$ respectively, which is very
consistent with strict Lorentz Invariance up to the highest energy.

\end{abstract}
\maketitle

\section{Introduction}

A number of extensions of the standard model suggest that Lorentz
Invariance (LI) is only a low-energy approximation and it may be
deformed at very high energies, e.g., approaching the Planck Scale
$\sim 10^{28}$ eV
\cite{Colladay:1996iz,Sarkar:2002mg,Stecker:2004pg,Li:2007vz}.
In a simple form shown by Coleman
and Glashow, Lorentz Invariance Violation (LIV) can be expressed
as a modified energy-momentum relation $E^2=m^2+p^2+\xi p^2$,
under the assumption that LI is violated perturbatively in the context
of conventional quantum field theory \cite{Coleman:1998ti}.
It can also be interpreted in terms of different maximal attainable
velocities for different particles. Modified energy-momentum relation
with LIV terms proportional to the cube of the momentum, or a higher
power, are also considered \cite{Morgan:2007dj}. Although LI has been
confirmed at accelerators up to $2$ TeV for protons\cite{Kahniashvili:2008va},
104.5 GeV for electrons and 300 GeV for photons\cite{Hohensee:2008xz},
it is still possible to see LIV in astrophysical processes with much
higher energy, especially in the Ultra High Energy Cosmic Rays (UHECRs)
with energies above $10^{18}$ eV \cite{Stecker:2004xm}.

Since the gyration radius of UHECRs is larger than the height of our
Galaxy in the Galactic magnetic field, UHECRs are generally thought
to be of extragalactic origin. Propagating through intergalactic
space, the UHECRs will interact with the Cosmic Microwave Background
(CMB) photons, which results in energy and flux depletion. In particular, the
photomeson process will induce a suppression in the spectrum above
$(3-6)\times 10^{19}$ eV and lead to the well-known 
Greisen-Zatsepin-Kuzmin (GZK) cutoff \cite{Greisen:1966jv,Zatsepin:1966jv}.
The spectrum of UHECRs can be calculated theoretically by assuming
the source distribution and injection energy spectrum. The energy loss
processes for UHECRs propagating in the intergalactic space include
photopion production and $e^+e^-$ pair production when interacting
with CMB, as well as the adiabatic energy loss due to the expansion
of the Universe \cite{Scully:2000dr,Berezinsky:1988wi,
Berezinsky:2002nc}. Because of the short mean-free path of photopion
production, the spectrum of UHECRs above $6\times 10^{19}$ eV falls
sharply and results in the GZK cutoff.

However, measurements of the UHECR spectrum have led to great confusion
in the last decade. The result of the Akeno-AGASA experiment clearly shows an
extension of the spectrum beyond the GZK cutoff
\cite{Shinozaki:2004nh}. To account for the AGASA data beyond the
GZK cutoff, LIV has been introduced \cite{Coleman:1998ti}. Even a
LIV parameter as small as $\xi \approx 3 \times 10^{-23}$ may lead
to the removal of GZK cutoff\cite{Coleman:1998ti,Stecker:2004xm,Bietenholz:2008ni}.
However, the measurements by HiRes \cite{Abbasi:2002ta} and Yakutsk
\cite{Egorova:2004cm} seem to show the existence of the GZK cutoff.

Recently, the situation tends to be clear. Having accumulated
data for years, the HiRes Collaboration confirms their previous
result and observes the GZK cutoff with a $5\sigma$ standard deviation
\cite{Abbasi:2007sv}. The Pierre Auger Collaboration gives results consistent
with HiRes and rejects a single power-law spectrum above $\sim 10^{19}$eV
at the $6\sigma$ confidence level \cite{Yamamoto:2007xj,Abraham:2008ru}.
As confirmation of the GZK cutoff, severe constraints can be placed
on the effect of LIV.

In this paper, we investigate how the LIV can be constrained
according to the latest HiRes and Auger data.
We give details of the method to calculate the UHECR spectrum with
LI in Sec. II. Then LIV is introduced to our calculation of the
spectrum of UHECRs in Sec. III. The modified spectrum with different
LIV parameters and the constraints on the LIV parameters are also shown
in this section. Section IV gives conclusions and discussions.

\section{The Spectrum of Ultra High Energy Protons in the Standard Model}

We assume the composition of UHECRs is
pure proton. When propagating in intergalactic space,
the ultra-high energy (UHE) protons will experience energy losses
through the adiabatic expansion of the Universe, $e^+e^-$ pair
production, and photopion production due to the interaction with CMB photons.
Then the energy evolution equation for a proton is
\begin{equation}
\label{energy}
-\frac{1}{E}\frac{dE}{dt}=\beta^{ad}_z(E)+\beta^{e^+e^-}_z(E)+
\beta^{\pi}_z(E)\ ,
\end{equation}
where $\beta^{ad}_z(E)$, $\beta^{e^+e^-}_z(E)$ and $\beta^{\pi}_z(E)$
are the proton energy loss rate due to the Universe expansion, pair
production and pion production respectively.

The energy loss rates of protons at $z=0$ are
\cite{Stecker:1968uc,Berezinsky:2002nc}
\begin{eqnarray}
\beta_0^{\rm ad}(E)&=&H_0,\nonumber\\
\beta_0^{e^+e^-,\ \pi}(E) & = &
\frac{1}{2 \gamma^2} \int^\infty_{\epsilon'_{\rm th}} \sigma(\epsilon')
K(\epsilon') \epsilon' {\rm d}\epsilon' \int_\frac{\epsilon'}
{2\gamma}^\infty \frac{n(\epsilon)}{\epsilon^2} {\rm d}\epsilon \nonumber\\
& = & \frac{T}{2 \pi^2 \gamma^2} \int^\infty_{\epsilon'_{\rm th}}
{\rm d}\epsilon' \sigma(\epsilon') K(\epsilon') \epsilon' \left\{-\ln
\left[1-\exp\left(-\frac{\epsilon'}{2\gamma T}\right)\right]\right\},
\label{beta}
\end{eqnarray}
where $H_0$ is today's Hubble expansion rate, $E$ and $\epsilon$ are
the energies of the proton and the CMB photon in the laboratory system (LS),
respectively, $\epsilon'$ is the photon energy in the proton rest system,
$\gamma$ is the Lorentz factor of the proton in the LS,
$\sigma(\epsilon')$ is the interaction cross section, $K(\epsilon')$ is
the average fraction of energy loss, i.e., the inelasticity in the LS,
$n(\epsilon)$ is the differential number density of CMB photons and
$T\approx 2.73$ K is the temperature of CMB. $\epsilon'_{\rm th}$ in
Eq.(\ref{beta}) is the threshold energy of the photon in the
proton rest system above which the $e^+e^-$ pair or pion production
can occur. Thus,
$\epsilon^{\prime e^+e^-}_{\rm th}=2m_e(1+m_e/m_p) = 1.022$ MeV
and $\epsilon^{\prime\pi}_{\rm th}=m_{\pi}(1+m_{\pi}/2m_p) = 149$ MeV
for $p\gamma \to pe^+e^-$ and $p\gamma  \to p\pi $ respectively.

Because of the redshifts of CMB photons, the energy loss rate will
be larger at redshift $z$. The energy loss rate $\beta_z(E)$ can be
derived as \cite{Berezinsky:2002nc}
\begin{equation}
\beta_z^{\rm ad}(E)=H(z),\ \
\beta_z^{e^+e^-,\pi}(E)=(1+z)^3\beta_0[(1+z)E],
\end{equation}
where $H(z)=H_0\sqrt{\Omega_M(1+z)^3+\Omega_{\Lambda}}$ is the Hubble
parameter. In this work we use the following cosmological parameters
$H_0=71$ km s$^{-1}$ Mpc$^{-1}$, $\Omega_M=0.27$ and $\Omega_{\Lambda}=
0.73$ according to the recent observations \cite{Komatsu:2008hk}.

Solving Eq. (\ref{energy}) with the boundary condition $E(z=0)=E_0$, we can get
the initial energy distribution of protons which will be observed with
energy $E_0$ at the Earth. We denote this function as $E_g(E_0,z)$, which
means the initial energy distribution as a function of redshift
$z$ and observational energy $E_0$. We employ two assumptions: (1)
proton sources are distributed homogeneously in the Universe without
the evolution effect; (2) the source spectrum is a power-law with index
$\gamma_g$. Then the observational proton spectrum at the Earth can
be written as \cite{Berezinsky:2002nc}
\begin{equation}
J(E_0)=\frac{L_0}{4\pi}(\gamma_g-2)\int_0^{z_{\rm max}(E_0)} {\rm d}z
\left|\frac{{\rm d}t}{{\rm d}z} \right|(1+z)^m E_g^{-\gamma_g}\frac{{\rm d}
E_{g}(E_0,z)}{{\rm d}E_0},\label{j0}
\end{equation}
with
\begin{equation}
\frac{{\rm d}E_{g}(E_0,z)}{{\rm d}E_0}=(1+z)\exp \left[\frac{1}{H_0}
\int_0^z {\rm d}z' \frac{(1+z')^2}{\sqrt{\Omega_m(1+z')^3+\Omega_{\Lambda}}}
\left(\frac{{\rm d}b_0(E')}{{\rm d}E'}\right)_{E'=(1+z')E_{g}(E_0,z')}\right],
\end{equation}
where $b_0(E)\equiv -{\rm d}E/{\rm d}t=E\beta_0(E)$, $|{\rm d}t/{\rm d}z|=
1/[(1+z)H(z)]$,
and $z_{\rm max}$ is the redshift of protons with maximum energy
$E_{\rm max}$ that reach us with energy $E_0$. In this work we adopt
$E_{\rm max}\approx 10^{22}$ eV. Larger $E_{\rm max}$ do not affect
the results. $L_0$ is the total luminosity of UHE protons and is determined
by matching the calculated spectrum to the observational data.
$(1+z)^m$ indicates the evolution of primary UHECR sources.
However, this term is still unclear at present.
Among the possible sources of UHECRs,
the galaxies and some types of active galactic nuclei 
show $m \approx 2.6$ redshift evolution
in radio, optical and X-ray bands\cite{Ueda:2003yx},
while for BL Lacs there is a strong ``negative'' evolution\cite{Morris:1991}.
It can be proven that for several reasonable evolution regimes,
including the no-evolution case,
the UHECR spectra can be reproduced well with different primary
spectra\cite{Berezinsky:2002nc}.
Therefore, we adopt the no-evolution case for 
the source luminosity($m=0$) in this work.

Eq.(\ref{j0}) can be simply understood: the protons within the initial
energy interval $(E_g,E_g+{\rm d}E_g)$ at redshift $z$ contribute to
the detected energy interval $(E_0,E_0+{\rm d}E_0)$; the sum of all
redshifts gives the total flux.

\begin{figure}[!htb]
\begin{center}
\includegraphics[width=8cm]{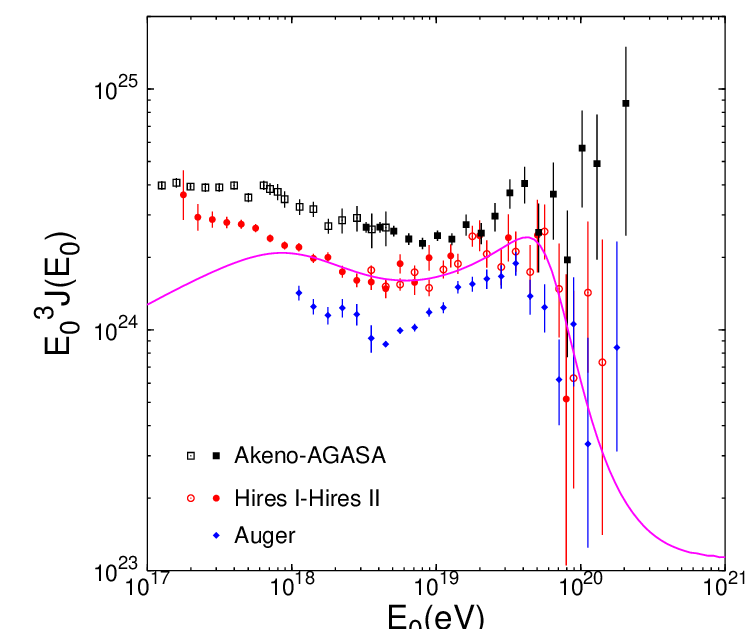}
\includegraphics[width=8cm]{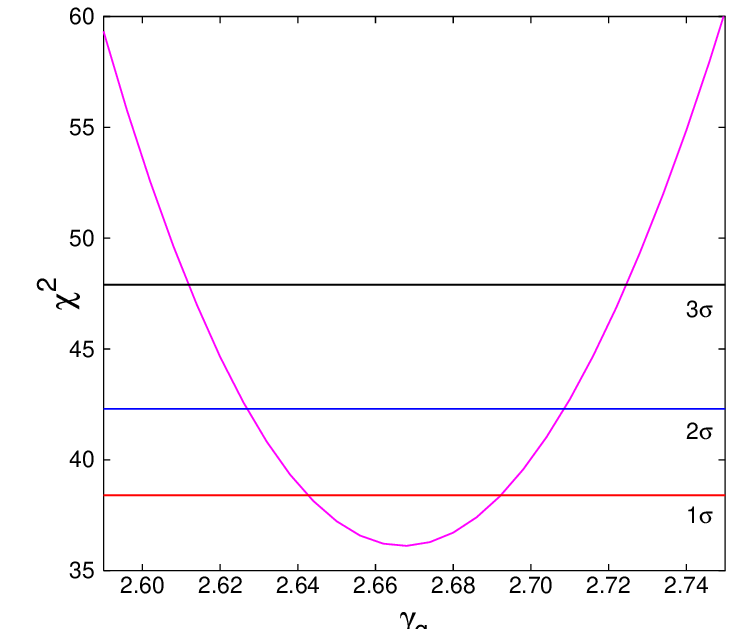}
\caption{Left panel: the expected UHE proton energy spectrum (magenta curve)
compared with the observational data from the AGASA \cite{Shinozaki:2004nh},
HiRes \cite{Abbasi:2007sv}, and Auger \cite{Yamamoto:2007xj} experiments.
The source spectrum of the theoretical curve is adopted as
$\gamma_g=2.67$, and the flux is normalized to the data of HiRes above
$10^{18}$ eV. Right panel: the fitting $\chi^2$ distribution as a function
of the parameter $\gamma_g$ for HiRes data. The three horizontal lines show the
uncertainty ranges of $\gamma_g$ at $1\sigma$, $2\sigma$ and $3\sigma$ 
confidence levels, respectively.}
\label{fig1}
\end{center}
\end{figure}

In Fig. \ref{fig1} we show the theoretical spectrum of the UHE protons
with the source index $\gamma_g=2.67$, which corresponds to the best fitting
results of HiRes monocular data \cite{Abbasi:2007sv}. Data from AGASA
\cite{Shinozaki:2004nh} and Auger \cite{Yamamoto:2007xj} are also
shown in Fig. \ref{fig1}. The theoretical flux has been normalized to
the data of HiRes above $10^{18}$ eV.
The cosmic rays below $10^{18}$ eV are usually thought to be of Galactic origin
\cite{Aloisio:2006wv}, and we ignore them in this work.
The right panel of Fig.\ref{fig1} shows the fitting $\chi^2$
distribution as a function of the source spectrum index $\gamma_g$ using
the HiRes data. The best fitting $\gamma_g$ is $2.67$ , with a $\chi^2/
{\rm d.o.f.}=36.1/31$. The statistical $1\sigma$ range of $\gamma_g$ is
$2.58<\gamma_g<2.73$.

\section{The Spectrum of UHE Protons with LIV}

In this section we discuss how the LIV can modify the UHE proton spectrum.
We adopt the framework of LIV developed by Coleman and Glashow,
that a small first order perturbation is added to the free particle
Lagrangian \cite{Coleman:1998ti}. The modification of the Lagrangian is then
translated into the change of the dispersion relation of free particles,
\begin{equation}
E^2=m^2+p^2+\xi p^2,
\label{dis}
\end{equation}
where $\xi p^2$ is the perturbation term. $|\xi|$ is a very small parameter
($\sim 10^{-23}$) and may be different for various particle species.

Since the dispersion relation is changed by LIV, the kinematics of the
reaction $p\gamma\to NX$ with $N$ the nuclei and $X$ the mesons, will
also be changed, and finally, the UHE proton spectrum is modified.
For a detailed treatment of the effect of LIV on the kinematics of
$p\gamma$ collision, we will closely follow the work of Alfaro and Palma
\cite{Alfaro:2002ya}. In this work we only consider the process
$p\gamma\to N\pi$, for simplicity.

The LIV effect generally modifies the kinematical inelasticity $K$,
which is defined as the ratio between the energy of pion, $E_\pi$,
and the primary proton energy $E_p$, $K=E_\pi/E_p$. The
equation of inelasticity
under the modified kinematics is given as \cite{Alfaro:2002ya}
\begin{equation}
(1-K_{\theta})\sqrt{s}= F
+\beta\cos\theta \sqrt{ F^2 -
s_N(K_{\theta})}\ ,
\label{ktheta}
\end{equation}
with
\begin{equation}
F=\frac{s+s_N(K_{\theta})-s_{\pi}(K_\theta)}{2\sqrt{s}}\ ,
\end{equation}
where $K_{\theta}$ is the inelasticity as a function of the angle
$\theta$ between the proton momentum in the center of mass system (CMS)
and the direction of the CMS relative to the LS, $s=E^2_{\rm tot}-
p^2_{\rm tot}=(E+\epsilon)^2-(\vec{p}-\vec{k})^2$ is the square of
the total rest energy in the CMS, and $s_\pi$ and $s_N$ are the CMS
energies of the pion and recoil nuclei, defined as
\begin{equation}
s_a=E^2_a-p^2_a=m_a^2+\xi_ap_a^2.
\end{equation}
Note that for $\xi<0$ $s_a$ can be negative and the particle
will have a spacelike four-momentum. We require $s_a>0$ in this
work to guarantee that the particle is timelike \cite{Alfaro:2002ya}.
According to the definition of inelasticity, we have
\begin{eqnarray}
s_N&=& m_N^2+\xi_N[(1-K_\theta)E_p]^2,\nonumber \\
s_\pi&=&m_{\pi}^2+\xi_\pi(K_\theta E_p)^2,
\end{eqnarray}
in which we replace the perturbation term $\xi p^2$ in Eq.(\ref{dis})
with $\xi E^2$, for simplicity.

Following Ref. \cite{Alfaro:2002ya} we assume $|\xi_\pi|\gg|\xi_N|\approx 0$,
and denote $\xi_\pi$ as $\xi$, then Eq.(\ref{ktheta}) can be further
simplified. Eq.(\ref{ktheta}) is solved numerically to get $K_\theta$,
and the average with respect to $\theta$ gives the total inelasticity
$K=\frac{1}{\pi}\int_0^{\pi}K_{\theta}{\rm d}\theta$.
We can see from the above equations that the inelasticity $K$ is a function
of $E_p$ and $s=(E+\epsilon)^2-(\vec{p}-\vec{k})^2=s_p+2\sqrt{s_p}\epsilon'$,
where $s_p$ is the energy of the proton in its CMS and $\epsilon'$ is the
photon energy in this system. For LI with $\xi=0$, $K$ is independent of
$E_p$ and is only a function of $\epsilon'$.

\begin{figure}[!htb]
\begin{center}
\includegraphics[width=10cm]{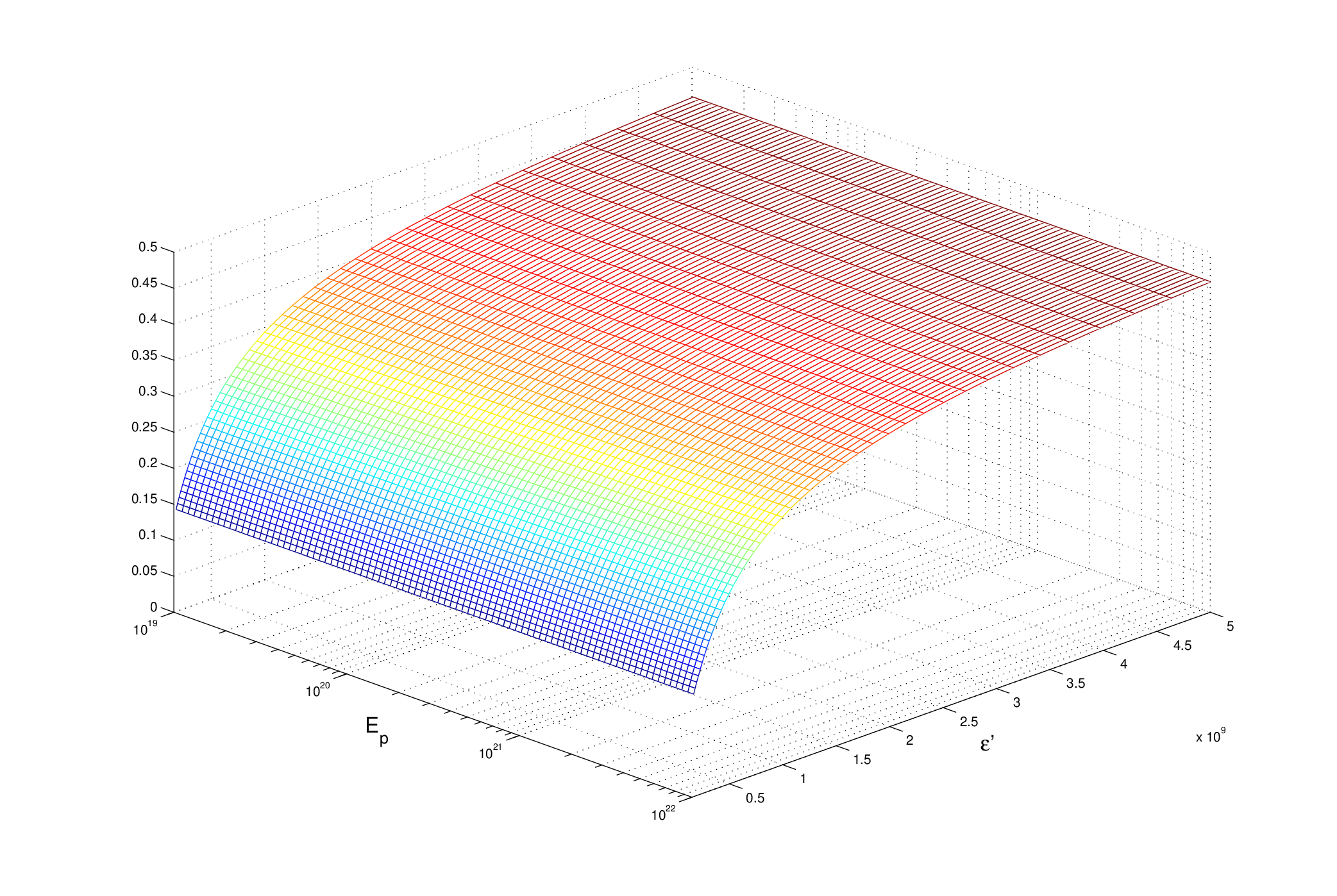}
\includegraphics[width=10cm]{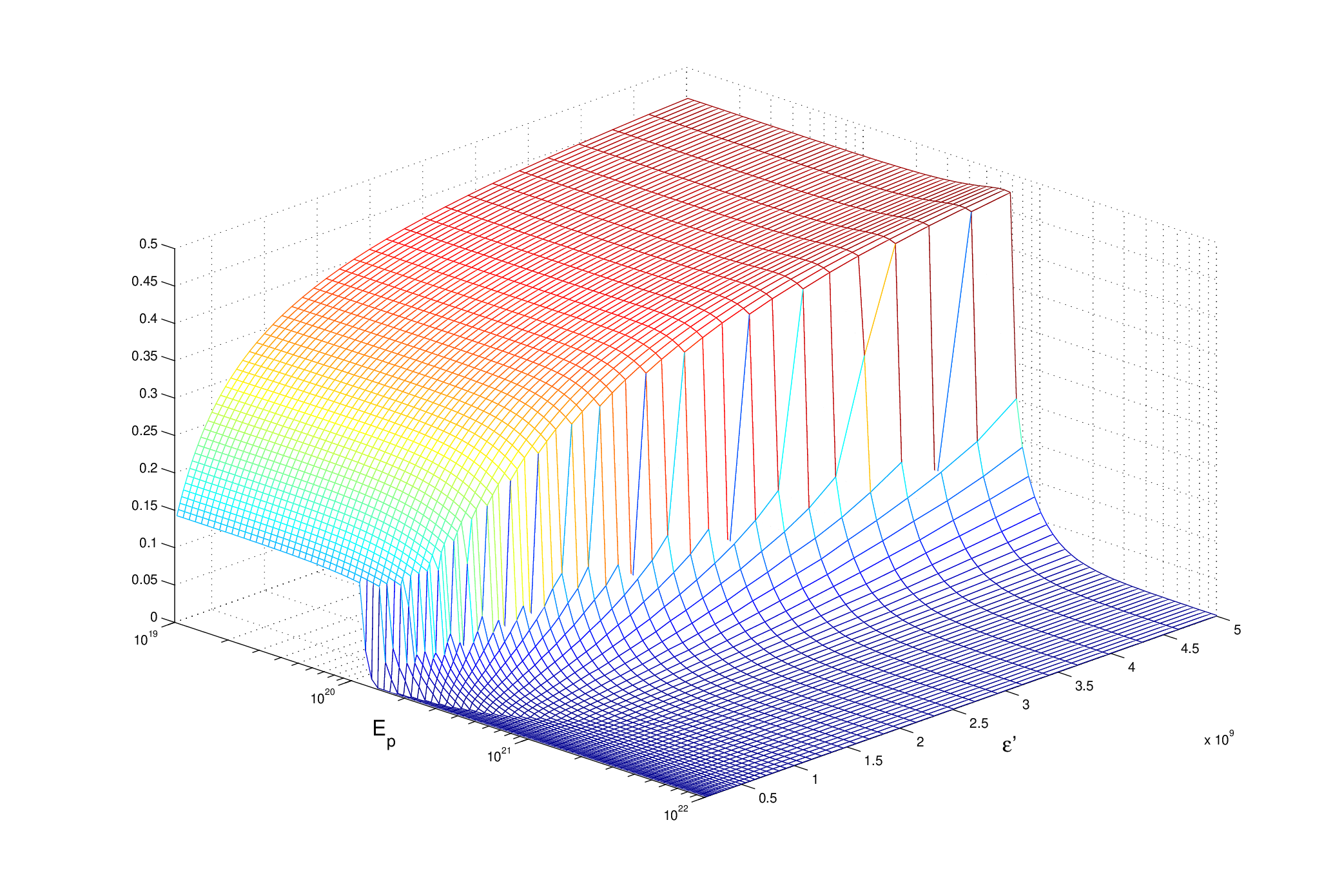}
\includegraphics[width=10cm]{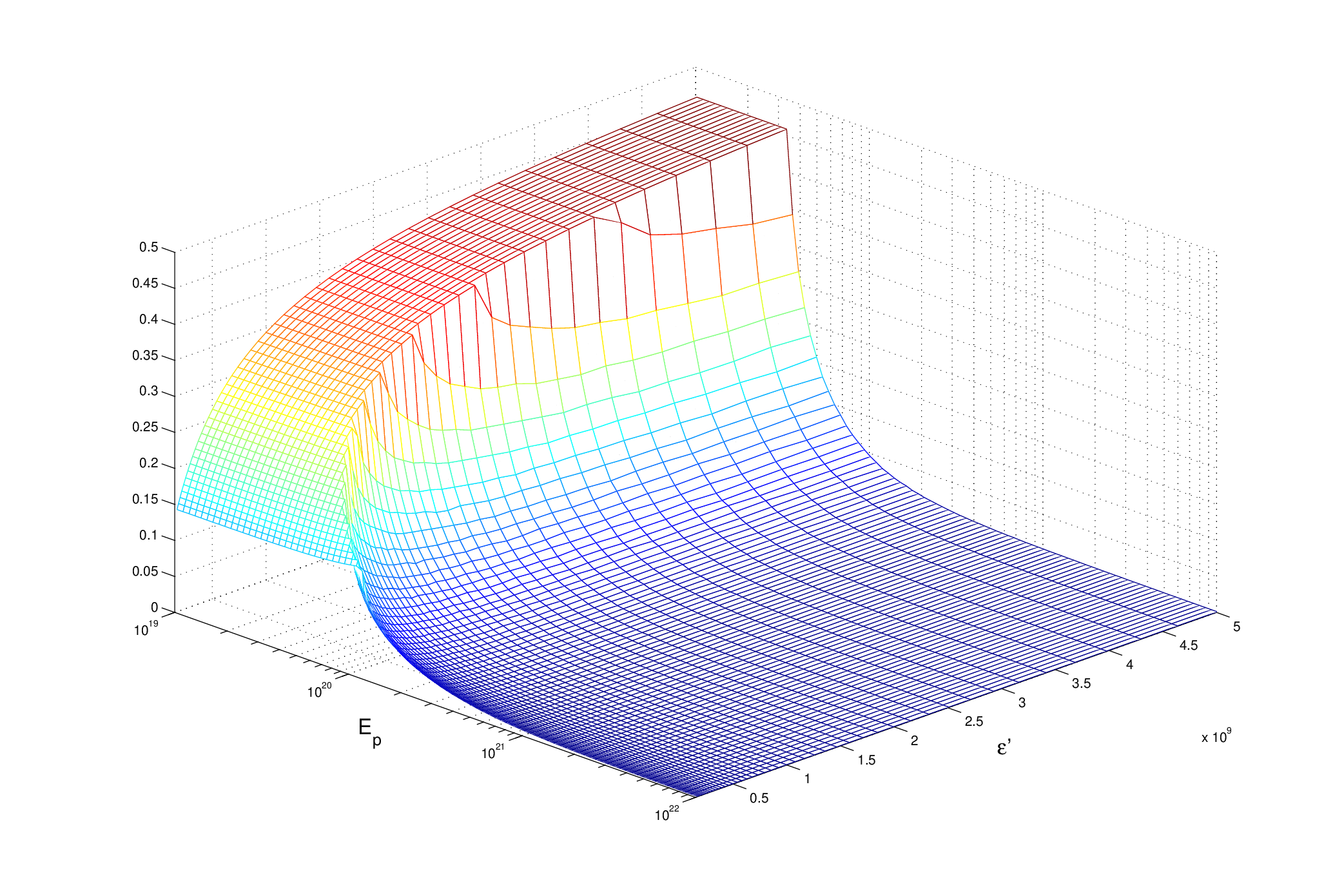}
\caption{The inelasticity as a function of the proton energy $E_p$ in
the LS and the CMB photon energy $\epsilon'$ in the proton rest system,
for the standard model (upper panel), LIV modified models with
$\xi=1 \times 10^{-23}$ (middle panel), and $\xi=-1 \times 10^{-23}$
(lower panel), respectively.}
\label{fig2}
\end{center}
\end{figure}

\begin{figure}[!thb]
\begin{center}
\includegraphics[width=8cm]{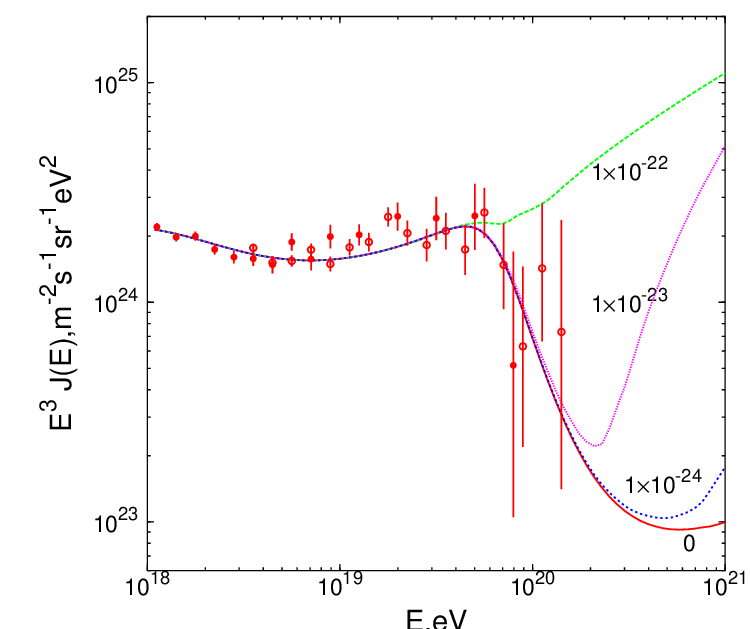}
\includegraphics[width=8cm]{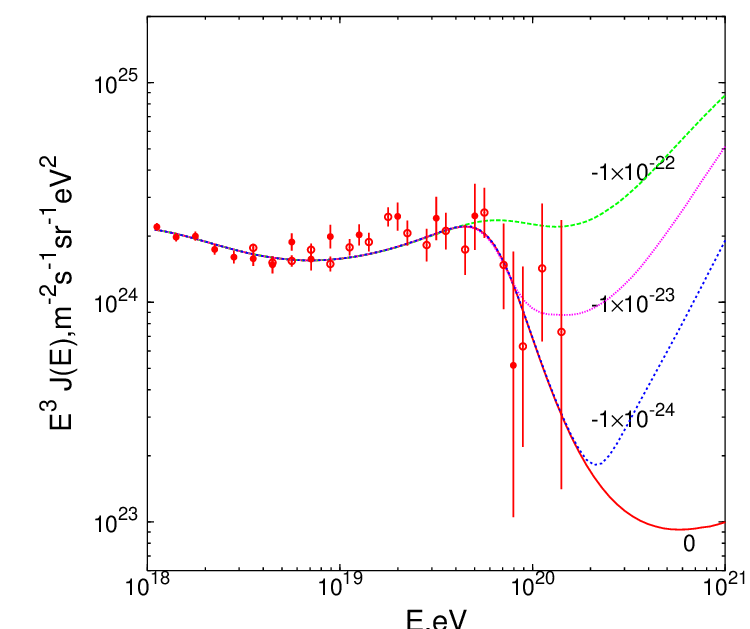}
\caption{The UHE proton spectra modified by LIV compared with 
the one predicted by the standard
model. Left panel: $\xi>0$; right panel: $\xi<0$.
The observational data are from HiRes \cite{Abbasi:2007sv}.}
\label{fig3}
\end{center}
\end{figure}

\begin{figure}[!htb]
\begin{center}
\includegraphics[width=8cm]{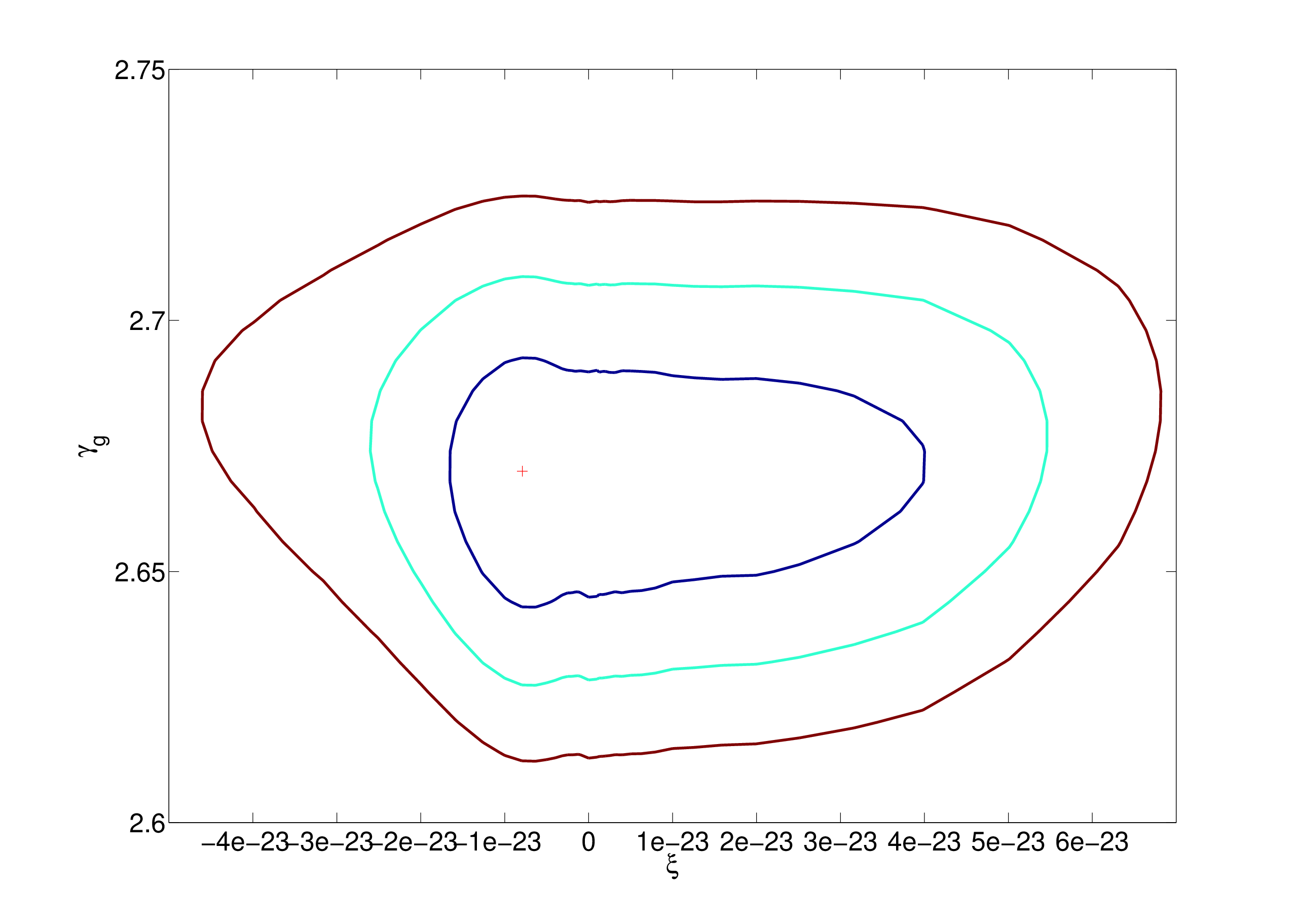}
\includegraphics[width=8cm]{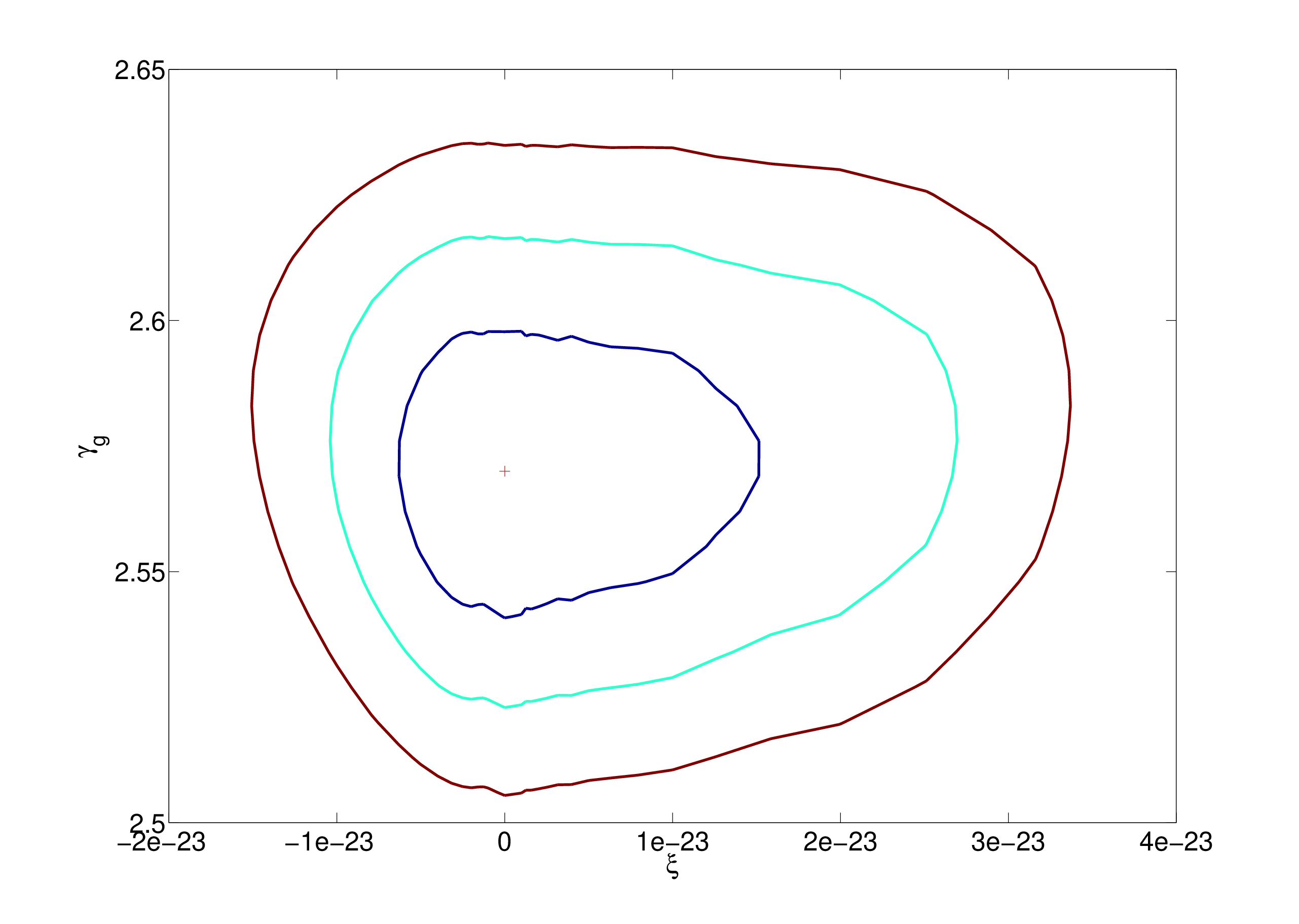}
\caption{The $1\sigma$, $2\sigma$ and $3\sigma$ confidence regions 
(from inside to outside) on the $\xi-\gamma_g$ parameter plane using HiRes
(left panel \cite{Abbasi:2007sv}) and Auger (right panel \cite{Yamamoto:2007xj})
data above $10^{18}$eV. The red cross shows the best fitting values.}
\label{fig4}
\end{center}
\end{figure}

In Fig. \ref{fig2} we show the inelasticity as a function of the proton
energy $E_p$ in the LS and the CMB photon energy $\epsilon'$ in the
proton rest system, for the standard model as well as the models modified
by LIV. Two kinds of LIV scenarios with $\xi=1 \times 10^{-23}$
and $\xi=-1 \times 10^{-23}$ are
investigated. We show that the modification of LIV is the reduction of the
inelasticity at high energies. This effect is understood as
the LIV leading to a reduction of the allowed phase space for
the interaction \cite{Alfaro:2002ya,Scully:2008jp}. It is interesting
to note that no matter if $\xi$ is positive or negative, the effect
is similar---the inelasticity is reduced. Therefore, if the LIV exists,
we can expect that the spectrum of UHE protons will be less suppressed
by the $\gamma p$ interaction.

The modified spectra of UHE protons for several values of LIV parameters
are shown in Fig. \ref{fig3}, together with
the unmodified spectrum from the standard model. 
We see that the GZK suppression
effect becomes less significant for LIV cases. For very high energies or
for large magnitudes of LIV parameters, the source spectra tend not to be
distorted, i.e., the photopion production process $\gamma p\rightarrow N\pi$
does not play an important role any more.

We employ a minimum $\chi^2$ fitting method to derive the implication
of the HiRes and Auger data on the LIV parameter. A scan in the
$(\xi,\gamma_g)$ plane is taken to calculate the UHECR spectra and
the corresponding $\chi^2$. Minimizing the $\chi^2$ distribution,
we give a combined fit to get the source spectrum index $\gamma_g$
and the LIV parameter $\xi$ simultaneously.
The confidence regions for $1\sigma$, $2\sigma$ and  $3\sigma$ confidence levels
in the $\xi-\gamma_g$ plane are shown in Fig. \ref{fig4} for
the HiRes monocular spectra and the Auger combined spectrum, respectively. The best
fitting results and $1\sigma$ uncertainties of the parameters are
$\gamma_g=2.67^{+0.01}_{-0.02}$, $\xi=-0.8^{+3.2}_{-0.5}
\times10^{-23}$ for HiRes and $\gamma_g=2.57^{+0.02}_{-0.02}$,
$\xi=0.0^{+1.0}_{-0.4}\times10^{-23}$ for Auger respectively.
It is shown that the standard model with $\xi=0$ is very
consistent with the present data.
Compared to previous work, here we employ more strictly
statistical analysis and include the negative part of $\xi$.

Considering an evolution factor $(1+z)^{2.6}$,  we find that
the best fitting results are $\gamma_g=2.55^{+0.02}_{-0.01}$,
$\xi=-0.1^{+1.2}_{-1.2} \times10^{-23}$ for HiRes data, and
$\gamma_g=2.37^{+0.01}_{-0.02}$, $\xi=0.0^{+0.5}_{-0.5}\times10^{-23}$ for
Auger surface data respectively. It is shown that the source spectrum
differs a bit from the case with no source evolution. As for the LIV
parameter, the results are very consistent within the statistical errors,
and the conclusion is almost unchanged.

\section{Conclusion and Discussion}

The recent observations of the UHECR spectrum by HiRes \cite{Abbasi:2007sv}
and Auger \cite{Yamamoto:2007xj} show the existence of GZK suppression.
In this work we use these observational data to test the LIV
model and set constraint on the LIV parameter. The composition
of UHECRs is assumed to be pure proton. We then solve the propagation
equation of UHE protons in an expanding universe after incorporating
the LIV effect in the energy loss rate of protons. A minimum $\chi^2$
fit to the observational data above $10^{18}$ eV is adopted to derive
the source spectrum of UHE protons and the LIV parameter. We find that
the current data can limit the LIV parameter for pions to the level
$\sim 3\times 10^{-23}$. The standard model with the GZK cutoff is very
consistent with the observational data.

Only the LIV on the photopion production process is considered in this
work. The pair production occurs at lower energy and the LIV effect
might be less significant. To incorporate the LIV effect into the $e^+e^-$
production process, the analysis will be more complicated. In addition,
we need to restrict the treatment to the case $\xi_\pi \gg
\xi_N\approx 0$.
This has been shown to be due to the weakness
of the presentation of the theory. 
We are unable to determine whether the perturbation term comes from
the initial proton or the final state proton, which have different
energy in the LS \cite{Alfaro:2002ya}.

In this work the composition of UHECRs is assumed to be pure proton.
However, the composition of UHECRs is poorly known from the experimental
point of view. The observations of HiRes show that the UHECRs are proton
dominant \cite{Abbasi:2004nz}, while the results from the Auger experiment
indicate a mediate mass composition \cite{Unger:2007mc}. The determination
of the UHECR composition depends on the interaction model and has a relatively
large uncertainty at present. If the UHECRs are heavy nuclei dominant,
the main process during the propagation in the CMB photon field is
photo-disintegration, which will lead to the change of the composition
and energy of primary cosmic rays and, accordingly, form a GZK-like spectrum
\cite{Aloisio:2008pp,Aloisio:2008uc,Hooper:2008pm}.

We can make a rough estimate of the LIV effect in such a case, by
assuming the primary composition of UHECRs to be iron and
considering the process $^{56}Fe+\gamma(CMB)\rightarrow^{55}Mn+p$.
The analysis is greatly simplified by assuming $\xi_{Mn} \approx
\xi_p \approx 0$, while the physics is not changed as only the
difference of the $\xi$'s in the initial and final states is
relevant for the kinematics\cite{Coleman:1998ti}. On the one hand,
the interaction of photo-disintegration is possible only if
$m_{Fe}^2+4\epsilon E_{Fe}+\xi_{Fe}E_{Fe}^2\ge(m_{Mn}+m_p)^2$,
resulting in $\xi_{Fe}\ge-4\epsilon^2/[(m_{Mn}+m_p)^2-m_{Fe}^2]
\approx-2\times10^{-25}[\epsilon/\epsilon_0]^2$ for CMB energy
$\epsilon_0=2.35\times10^{-4}$eV. On the other hand, the
spontaneous fragmentation $^{56}Fe\rightarrow^{56}Mn+p$ should be
forbidden for iron with energy lower than $\sim 3\times10^{20}$eV
since cosmic rays with such energies have been detected. This
condition requires $m_{Fe}^2+\xi_{Fe}E_{Fe}^2<(m_{Mn}+m_p)^2$,
which gives $\xi_{Fe}<1.2\times10^{-23}$ for
$E_{Fe}=3\times10^{20}$eV. Therefore, we get
$-2\times10^{-25}\le \xi_{Fe} \le 1.2\times10^{-23}$. It
should be noted that the above estimates are quite rough. It will
be more complicated to use the UHECR spectrum to study the LIV
effects for heavy nuclei because there will be a chain of nuclei
species taking effect in the interactions. Further detailed
analysis is needed to probe the LIV if the UHECRs are proved to be
heavy nuclei in future experiments.

\acknowledgments{We are grateful to Veniamin Berezinsky for helpful correspondence.
This work is supported by the NSF of China under Grant
No. 10575111 and No. 10773011, and in part by the Chinese Academy of
Sciences under Grant No. KJCX3-SYW-N2.
}


\end{document}